\documentclass{optica-article}
\journal{opticajournal} % for journals or Optica Open
\articletype{Research Article}
\usepackage{xcolor}

\begin{document}

\title{%Exciton-polaritons as a platform for digital computing, neuromorphic computing and optimization
Harnessing exciton-polaritons for digital computing, neuromorphic computing, and optimization}

\author{Andrzej Opala,\authormark{1,2} Michał Matuszewski,\authormark{2,3,*}}

\address{\authormark{1}Institute of Experimental Physics, Faculty of Physics,\\University of Warsaw, ul. Pasteura 5, PL-02-093 Warsaw, Poland\\
\authormark{2}Institute of Physics, Polish Academy of Sciences,\\ Aleja Lotnik\'ow 32/46, PL-02-668 Warsaw, Poland\\
\authormark{3}Center for Theoretical Physics, Polish Academy of Sciences,\\ Aleja Lotnik\'ow 32/46, PL-02-668 Warsaw, Poland}

\email{\authormark{*}mmatu@ifpan.edu.pl} 

\begin{abstract*} 
Polaritons are quasiparticles resulting from strong quantum coupling of light and matter. Peculiar properties of polaritons are a mixture of physics usually restricted to one of these realms, making them interesting for study not only from the fundamental point of view but also for applications. In recent years, many studies have been devoted to the potential use of exciton-polaritons for computing. Very recently, it has been shown experimentally that they can be harnessed not only for digital computing, but also for optical neural networks and for optimization related to hard computational problems. Here, we provide a brief review of recent studies and most important results in this area. We focus our attention in particular on the emerging concepts of non-von-Neumann computing schemes and their realizations in exciton-polariton systems.
\end{abstract*}

\section{Introduction}

Computation in the modern world is dominated by electronics, where data processing is realized with electronic currents. On the other hand, optics is used for efficient communication, in particular over long distances and at high data rates, as in long-haul optical fiber connections. This dichotomy stems from physical properties of light and matter. Material particles such as electrons are characterized by considerable interparticle interactions, which results in nonlinear input-output characteristics. This makes it easy to construct information processing elements that require nonlinearity, such as transistors, switches, memory cells etc. In contrast, according to Maxwell equations, photons do not interact in vacuum.  On the other hand, propagation of optical signals at the speed of light and at very small dissipation rates makes light ideal for use in communications.

The use of optical fields for efficient computations has been a long standing dream. Although quantum field theory predicts nonlinear behaviour of electromagnetic fields in vacuum at extremely high intensities~\cite{Bialynicki-Birula_1988}, the deviation from linearity is negligible in practice. When light interacts with matter, nonlinear effects of optical fields appear at realistic field intensities~\cite{boyd2020nonlinear}. However, these are in practice much weaker than the nonlinearity present in electronic systems. In most cases they require very high field intensities, which limits their usability. Moreover, in real optical devices it is difficult to fulfill the requirements for the operation of a complete large-scale device, such as cascadability, fan-out or logic level restoration~\cite{miller2010optical}.

Polaritons are hybrid quantum quasiparticles in which matter excitations are dressed with photons. They come in a variety of forms depending on the kind of matter particles that couples to light and the form of the coupling. Microcavity exciton-polaritons, where photons enclosed in a cavity couple to excitons, are among the most interesting and most intensively investigated types of polaritons. They are typically observed in semiconductor heterostructures where semiconductor layers of different molecular composition form both mirrors and the cavity sandwiched between them. A number of quantum wells hosting excitons can be inserted in the cavity, in the positions such that the coupling to the cavity mode is enhanced. This configuration allows to achieve very high light-matter coupling at very low dissipation rates.  Exciton-polaritons possess peculiar physical properites, which allowed to observe a range of phenomena including bosonic condensation~\cite{kasprzak2006bose}, effects of superfluidity~\cite{amo2009superfluidity,lerario2017room}, quantum vortices~\cite{nardin2011hydrodynamic,sanvitto2011all}, non-Hermitian physics~\cite{gao2015observation}, and topological states~\cite{klembt2018exciton}. Exciton-polaritons are also promising for practical devices working at room temperature~\cite{sanvitto2016road}.  

The aim of this review is to provide a brief overview of the research on exciton-polaritons for computation and optimization. This topic attracted strong interest in recent years due to their physical properties. Importantly, the interparticle interaction between polaritons can be orders of magnitude stronger than between photons~\cite{Matuszewski_2021} thanks to the strong quantum coupling with the matter component. On the other hand, their excellent transport properties are inherited from the light component. Polaritons can be excited simply by using external laser light incident on an appropriately designed optical cavity, if losses and the decoherence rate are weaker than the light-matter coupling strength. In other words, dressing light with matter can give photons a "material" boost that dramatically affects their physical properties. 

The review is divided into parts, each devoted to a different approach to using polaritons in computing. In Sec.~\ref{sec:digital} we summarize theoretical and experimental results on using polaritons for realizing standard computing operations based on digital logic, which mimic common electronic chips. These operations can be essential parts of optical machines that fit into the von-Neumann computational paradigm, with separate processing and memory units. In Sec.~\ref{sec:neuromorphic} we provide a short introduction to the concept of neuromorphic computing, and summarize results obtained in the context of possible applications of exciton-polaritons. In contrast to the standard digital approach, polariton neural networks do not require fully functional gates, transistors or memory cells, if the physical system itself is designed to resemble the neural network model. Finally, in Sec.~\ref{sec:optimization} we discuss yet another promising approach to computing, which is based on the experimental realization of optimization related to certain known hard problems by appropriately constructed polariton networks. In prinicple, at a sufficiently large scale, such specialized optimization devices would be able to solve problems that are beyond the capability of any classical algorithm. In result, they could efficiently solve problems that are intractable for classical computers working in the von Neumann architecture~\cite{stroev2023renaissance}. Finally, in Sec.~\ref{sec:other} we discuss other emerging approaches to polariton-based computing, including quantum computing and quantum neural networks.

\section{Digital computing with polaritons} \label{sec:digital}

\subsection{Polariton-based optical switches}

The sensivity of exciton-polaritons to electrical, optical, and other fields combined with high propagation velocities inspired research into digital computing with polaritons. Polaritonic analogs of basic elements of digital computing such as optical or electrooptical switches, transistors, logic gates have been constructed. The possibility to realize a sensitive polaritonic switch in the regime of optical parameteric oscillation was considered already in~\cite{zhang2007phase}. The first experimental realization of an exciton-polariton all-optical switch involved a wide sub-threshold pump laser slightly detuned from resonance and a tighly localized probe that provided the extra light intensity~\cite{amo2010exciton}.%, see Fig.~\ref{fig:fig_1}(a). 
The energy blueshift of the polariton state induced by the probe resulted in resonant transmission in the entire pump area, leading to a dramatic change in transmission.  In additon, the device was characterized by a high sensitivity to optical polarization of input beams. While this experiment was realized in the continuous wave regime, pulsed switching on timescales of the order of ten picoseconds were demonstrated in another work~\cite{steger2012single}. In~\cite{de2012control}, a configuration with two polariton fluids excited with two lasers with different frequencies was explored. It was shown that in the bistability regime, a single laser source was enough to control the hysteresis cycle. Optical control with picosecond pulses was also achieved.

A method of achieving polariton control by optical routing was proposed in~\cite{flayac2013exciton} and realized experimentally in~\cite{marsault2015realization}. This idea was based on a double barrier optical gate with periodically modulated waveguides on each side. The difference in modulation periods on the two sides resulted in asymmetric polariton flow from the nonresonantly pumped gate area, with the direction of the flow dependent on the pump intensity. Other proposed optical switching methods involved using nonresonant, polarized pulses to control the spin states of polariton condensates~\cite{li2015incoherent} and using the optical Stark effect to tune polariton energy~\cite{cancellieri2014ultrafast}. This effect was recently realized experimentally in the work~\cite{suarez2020electrically}, which also investigated picosecond effects of the so-called incoherent particle reservoir.

Switching of polariton fluids can be enhanced by appropriate shaping of the pump beam. This effect was demonstrated both in a multiple-spot pumping configuration and ring-pump geometry, where switching between complex spatial modes including petal and asymmetric modes was observed~\cite{cristofolini2013optical,sun2018stable}. Moreover, switching between stable magnetic polariton states was demonstrated~\cite{Ohadi2015Spontaneous}. In the latter case, optical control with 100-fold weaker resonant pulses of opposite spin was achieved.

Recent research on optical switching focuses on room temperature operation and more sophisticated phenomena. In 2021, single-photon control over polariton bosonic condensation was demonstrated in the experiment that involved a strong pump and a much weaker seed~\cite{zasedatelev2021single}. The experiment was carried out at room temperature in a microcavity filled with ladder-type polymer. This work opens the way to sub-attojoule switching and quantum optical polaritonic devices. All-optical switching at room temperature was also realized in perovskite microwires~\cite{feng2021all} and ZnO microcavity~\cite{chen2022optically}, where femtosecond swiching via condensate depletion was shown. More intricate observed phenomena include all-optical vortex switching~\cite{ma2020realization}, switching of topological phases~\cite{su2021optical}, and exploiting an exceptional point in a dissipative system for ultrasensitive switching~\cite{li2022switching}.

Apart from the optical switching discussed above, polaritons can be controlled by electrical and other fields to induce strongly nonlinear optical response~\cite{Liew2010electrical}. The remarkable result of~\cite{dreismann2016sub} shows that polaritons can be very sensitive to electrical stimuli, requiring as low as 0.5 fJ of electrical energy for spin switching.%, see Fig.~\ref{fig:fig_1}(b). 
Electrical control over polariton lasing was demonstrated in~\cite{suarez2020electrically} in a compact device that can be integrated into optical circuits. Another avenue for electrical control is the use of microcavities filled with liquid crystals and the associated spin-orbit coupling effect~\cite{rechcinska2019engineering}, as demonstrated in~\cite{lempicka2022electrical}. Electrooptical switching was also realized via polariton lasing in topological states~\cite{gagel2022electro}. Room-temperature electrooptic switches were realized in organic polariton microcavities~\cite{de2023room} and ferroelectric van der Waals semiconductors~\cite{luo2023electrically}. Moreover, it was shown experimentally that it is possible to manipulate polaritons  by strain fields~\cite{demenev2019ultrafast} and magnetic field~\cite{caputo2019magnetic}.

% \begin{figure}
% \centering
%   \includegraphics[width=\textwidth]{fig2.png}
%   \caption{{\bf Polaritons for digital computing.} (a) Switching-on of a near-resonant transmission through an exciton-polariton microcavity with a tightly focused probe beam~\cite{amo2010exciton}, where the scale indicates a distance of 25 $\mu$m. (b) Sub-femtojoule electrical switching of  circular polarization of a nonresonantly excited polariton condensate~\cite{dreismann2016sub}. (c) All-optical transistor based on a polariton flow along a microcavity ridge, gated by a weak blocking beam~\cite{gao2012polariton}. (d) Room-temperature optical transistor realized in a microcavity filled with an organic polymer~\cite{zasedatelev2019room} (e) Cascadability of polariton transistors based on an in-plane flow in a microcavity supported by resonant address and control beams~\cite{ballarini2013all}. }
% \label{fig:fig_1}
% \end{figure}

\subsection{Elements of polariton circuits}

The switching properties of polaritons make it possible to construct nonlinearly operating devices such as transistors, diodes, or logic gates. The first realization of a polariton logic gate, the XNOR gate, dates back to 2007~\cite{leyder2007interference,ostatnicky2010theory}. The experiment used the polarization sensitivity and interference between coherent polaritons for selective suppression of scattering.  It was also proposed that polariton bistability can be used for the construction of polariton propagation channels and logic gates~\cite{Shelykh_Neurons}.
An idea of a polariton spin transistor exploiting the polarization splitting was put forward in~\cite{shelykh2010optically}. A complete architecture for integrated photonic circuits based on spin-sensitive bistability was also proposed~\cite{espinosa2013complete}. Moreover, it was suggested that dissipative polariton solitons can be used for the implementation of logic gates~\cite{cancellieri2014ultrafast}.

Polariton transistors were realized experimentally  both in the case of excitation using nonresonant and resonant laser light. In the nonresonant case, a gate optical beam, much weaker than the pump beam, was able to block the polariton flow along a microcavity ridge~\cite{gao2012polariton}.%, see Fig.~\ref{fig:fig_1}(c). 
The dynamics of polaritons as well as their relaxation were investigated in detail~\cite{anton2012dynamics,Anton2013Relaxation}. In this configuration, the waves propagating in a waveguide were used to implement an AND logic gate~\cite{anton2013quantum}. Moreover, a polariton tunneling diode based on double-barrier resonant tunneling with optical nonresonant control beam was realized~\cite{nguyen2013realization}. However, the nonresonant excitation precludes full cascadability as the frequency of the output is necessarily lower than the frequency of the pump and control beams. Morevoer, energy of optical pump is to a large extent converted to the incoherent reservoir of excitons and uncondensed polaritons, which limits the potential efficiency of the device. To overcome this obstacle, a resonantly excited all-optical transistor was demonstrated, harnessing fluids propagating in the plane of a microcavity~\cite{ballarini2013all}.%, see Fig.~\ref{fig:fig_1}(e). 
In this configuration, amplification of the signal of up to 19 times  was achieved, with energy necessary for switching at the femtojoule level and switching times of the order of ten picoseconds. The first implementation of a resonantly pumped polariton transistor was also used for the realization of optical OR and AND logic gates. Directional transistor functionality in the optical parametric oscillation regime was also shown~\cite{lewandowski2017directional}. Polariton bistability was used in a prototype of an electrically-controlled rather than optically-controlled transistor~\cite{suchomel2017prototype}. The first room-temperature polariton transistor based on organic micorcavity was demonstrated in 2019, with the gain of 10dB$\mu$m$^{-1}$, and the demonstrated functionality of AND and OR gates~\cite{zasedatelev2019room}.%, see Fig.~\ref{fig:fig_1}(d). 
Moreover, a cascadable universal NOR gate was realized in an organic polariton microcavity using non-ground state dynamics~\cite{baranikov2020all}.

While elementary computation components such as logic gates and transistors are indispensable for the implementation of a von Neumann machine, it also requires other elements, including memory. This challenge was addressed in the polariton field by tristable ultrafast polariton memories~\cite{cerna2013ultrafast} and the proposal of vortex-based topological memories~\cite{sigurdsson2014information}. Moreover, information processing with exciton-polaritons can leverage their inherent properties, such as high in-plane velocity and long range propagation, as demonstrated in several works~\cite{Shelykh_Neurons,sigurdsson2015switching,lerario2017high,su2018room,luo2023electrically}.

%\newpage{}
%\clearpage{}
\section{Neuromorphic computing} \label{sec:neuromorphic}

\subsection{{Artificial neuron}}

The first mathematical models of artificial neurons were developed in the 40s~\cite{McCulloch_1943}. However, in the early research stage, these systems faced scepticism, which delayed progress in the discipline for almost four decades~\cite{Minsky_69}. In the 80s, new perspectives for artificial neurons and neural network applications appeared, giving momentum to the field. This progress was attributed to the pioneering works on optimisation algorithms~\cite{Rumelhart_1986} and rapid advancements in computing capabilities. Thanks to that, sophisticated neural networks, capable of processing large amounts of data and solving problems difficult for standard man-made algorithms, were created. Nowadays, neural networks became one of the primary approaches to machine learning and artificial intelligence, and are crucial for science and applications.  

The usefulness of neural networks is caused by their unique capability for generalisation and extracting correlations from external data. Generally, these computing systems are not designed to solve a specific problem given by algorithms but to learn how to do it. For this reason, artificial neural networks can be highly accurate in tasks in which traditional algorithms often fail, such as detection, classification and prediction.  To solve these problems accurately, the  number of neurons creating a neural network should be large enough. Nevertheless, even a single artificial neuron can solve simple machine learning tasks. 

The key to understanding how artificial neurons process data can be explained with a simple mathematical model. As in biological systems, an artificial neuron consists of multiple inputs and a core. The input information is multiplied by certain weights associated with each input. Next, the incoming rescaled signals are summed and directed to the neuron core. Signals reaching the neuron core can excite it according to the activation function $\varphi(\cdot)$. Mathematically this process can be expressed by the following equation
\begin{equation}
    y=\varphi{({\bf{W}}^T{\bf{x}}+b)=\varphi\bigg(\sum_{j=1}^{n}w_jx_j+b \bigg)},
\end{equation}
where $y$ is the neuron response, ${{\bf{W}}}^T=[w_1,w_2,w_3,\cdots, w_n]$ is the vector containing all synaptic weights, $\bf{x}$ is the vector of inputs ${\bf{x}}=[x_1,x_2,x_3,\cdots, x_n]$ and $b$ is the offset also called the bias. 

To illustrate the underlying concept explaining the computational capabilities of a system of artificial neurons, we consider the universal approximation theorem~\cite{Cybenko_1989}.
According to the postulates of the theorem proposed by Cybenko, a finite superposition of univariate and discriminatory functions can approximate any real n-variable continuous function $G(\mathbf{x})$. This theorem can be formulated as  
\begin{equation}
    G(\mathbf{x})=\sum^{N}_{j=1}\alpha_j\varphi(\mathbf{W}_j^T\mathbf{x}+b_j),
\end{equation}
where $N$  is the number of neurons, $\mathbf{x},\mathbf{W}\in \mathbb{R}^n$, while $\alpha_j,b_j \in \mathbb{R}$ are fixed parameters.  Hence, a system of sigmoid artificial neurons can be considered a universal function approximator. The universal approximation theorem can be generalised to a broad class of nonlinear functions, such as rectified linear unit (ReLU), Heaviside,  or hyperbolic tangent functions. Nowadays, the ReLU function is the most commonly used mathematical formula applied in artificial neuron implementations~\cite{Szandała_2021}. 

One can construct a complex computing system by connecting multiple artificial neurons. The computing performance of neural networks is not solely determined by the activation function, but also by the design of inter-neuron connections. The two main neural network types are feed-forward and recurrent neural networks.

\subsection{{Feed-forward neural networks}}

A feed-forward neural network (FFNN) is characterized by a layered architecture and unidirectional flow of information. Typically, connections of neurons are limited to neighbouring layers only, while neurons within the same layer process data independently. There are several exceptions to this rule, including residual neural networks, where connections between distant layers also exist. The input and output layers of a feed-forward network are directly responsible for collecting input data and returning results, respectively. For that reason, those layers are also called {\it{visible layers}}, in opposition to the layers between them, which are called {\it hidden layers}.  Usually, the neurons in hidden layers and the output layer can transform information in a nonlinear manner.

The simplest model of a feed-forward neural network contains two visible layers and one hidden layer. In this case, the equation describing how the neural network transforms the inputs $\bf{x}$ takes the following form
\begin{equation}
\label{eq:3}
    {\bf{y}}({\bf{x}})=f\big( {\bf{W}}_{\text{out}}\varphi\big( {\bf{W}}_{\text{in}}{\bf{x}+{\bf{b}}_{\text{in}}}\big)+{\bf{b}}_{\text{out}}            \big),
\end{equation}
where $f(\cdot)$ is the nonlinear function of the output layer, ${\bf{W_{\text{in}}}}$ and ${\bf{W_{\text{out}}}}$ are the matrices of weights containing connections between the hidden layer and the two visible layers. The network is called fully connected if all elements of the weight matrices can have non-zero values. The activation function of neurons in the hidden layer is $\varphi(\cdot)$. Vectors ${\bf{b}}_{\text{in}}$ and ${\bf{b}}_{\text{out}}$ are the biases of the input and output layers, respectively. 
%When the neural network parameters are chosen randomly, there is a negligible probability that the system described by equation (\ref{eq:3}) has the capability to solve effectively any complex tasks.  
During the training phase, parameters of the network (weights and biases) can be adjusted to obtain the optimal performance. Learning strategies can be categorised into three main paradigms: supervised, unsupervised and reinforcement learning. Here we focus our considerations on the supervised learning paradigm.

The supervised learning method requires a set of training examples, which are used to tune the network parameters to obtain optimal accuracy.  
This so-called training set is a collection of inputs $x_i$ collated with the desired neural network outputs $\mathcal{D}=\{(x_1,d_1),\dots,(x_m,d_m)\}$, where $m$ is the number of samples and $d_m$ are the targets. Weights and biases are tuned during training to achieve the lowest possible error rate between neural network outputs and target values. This error is quantified by the loss function, which can be defined, for instance, as the mean squared error
\begin{equation}
\label{eq:4}
L({\bf{W}}_{\text{in}},{\bf{W}}_{\text{out}},{\bf{b}}_{\text{in}},{\bf{b}}_{\text{out}})=\frac{1}{2}\sum_{j=1}^m\sum_{k=1}^{N_{\rm out}}\big(y_k^{(j)}-d_k^{(j)}\big)^2,
\end{equation}
 where $N_{\rm out}$ is the number of neurons in the output layer. 

 This procedure can be performed using the backpropagation algorithm, which is a widely employed and powerful approach to optimising the parameters of neural networks. In this algorithm, errors are calculated backwards through the network. The weights are tuned depending on the derivative of the loss function with respect to each parameter of the network. This process is performed for all examples from the training set to determine the global minimum of the loss function. The backpropagation algorithm can be applied to tackle very complex problems. Nevertheless, solving sophisticated problems using backpropagation algorithm applied to a deep multilayer neural network can be time consuming and computationally challenging. 
 
Another network model called the Extreme Learning Machine (ELM) was introduced to overcome these difficulties. Extreme learning machine is a feed-forward neural network with a single hidden layer and random connections between inputs and hidden layers~\cite{Huang_2006}. The concept of ELM was proposed in 2006 and successfully adapted to a wide range o machine learning tasks, including classification, regression, or creation of surrogate models and neuromorphic systems ~\cite{Pierangeli_2021}. What is unique in this network architecture is that in the training phase, only the output layer weights are modified. 

Therefore, the structure and number of neurons in the hidden layer play the main role in this kind of network model. Thanks to the randomness of neuron connections, inputs are transformed into a nontrivial multi-dimensional space in a nonlinear manner. The states in this space are used as sources of information extracted in the output layer in such a way that the task is performed optimally. It is worth mentioning that in some cases analytical formulas are used to determine optimal output weights.

\subsection{{Recurrent neural networks}}

In contrast to FFNN, recurrent neural networks (RNNs) have connections between neurons that can form loops, and operate on time-dependent data. Thanks to this recurrent structure, RNNs can manifest different types of memory, including short-term or long-term memory. RNNs are particularly efficient in computations where dynamics of the system plays a crucial role. 

As an example of a RNN, we consider a class of network models called reservoir neural networks. For clarity, we emphasize that reservoir neural networks should not be confused with the reservoir of incoherent particles appearing during condensation of exciton-polaritons. Neural networks that operate in the reservoir computing paradigm contain the visible layers and a single hidden layer called the {\it{reservoir}}~\cite{Jaeger_2004, Lukoševičius_2009, Tanaka_2019}, see Fig.~\ref{fig:fig_2}(a). As in the case of ELMs, the primary function of the reservoir is to map input signals into a higher-dimensional space of nonlinear reservoir nodes. This process can uncover hidden correlations between input signals that cannot be extracted by simple linear transformations. Reservoir networks can be used for classification or regression. Reservoir neural networks excel in prediction of nonlinear series and performing relatively simple algorithms. In a reservoir neural network, connections between neurons in the reservoir are usually random and can form feedback loops. Closed paths between neurons can lead to the emergence of highly nonlinear and chaotic dynamics of the reservoir. As in the case of an ELM, the weights of coupling between nodes in the hidden layer are non-tunable. All weights in the reservoir are fixed throughout the learning process, whereas only connections between the hidden and output layers are tuned. Importantly, even simple physical systems can be used as a reservoir. Hence, this computing architecture is commonly used in the hardware realisations of neural networks ~\cite{Larger_2012,Du_2017, Pierangeli_2020}.

%In contrast to feed-forward neural networks, the reservoir approach eschews hidden layers in favour of the reservoir containing numerous interconnected neurons. Establishing a feedback loop between reservoir neurons introduces highly nonlinear and chaotic dynamics. Additionally,  Owing to their recurrent nature, reservoir neural networks are particularly well-suited for machine-learning tasks, such as predicting nonlinear time series, dynamical control, and speech recognition. The increasing interest in reservoir neural networks is induced since the connections between nodes in the reservoir do not require modification during the teaching. 

Here we consider a simple rectangular lattice of $N \times N$ weakly coupled nonlinear oscillators as an example of a physical system which can be considered to be a reservoir. We assume that the coupling coefficients between nearest-neighbour lattice sites are given by $W_{nm}$ where $n$ goes through all lattice sites. The system can be modelled according to the following equation
\begin{equation}
\frac{d \psi_n}{d t}=\sum_{m=nn}{W_{nm}\psi_m}+f(\psi_n)\psi_n+u_n(t),
\end{equation}
where $\psi_n$ is the complex amplitude of the n-th oscillator. In the above equation, the first term on the right-hand side corresponds to the interaction between neighbouring lattice nodes. The second term contains a function $f(\psi_n)$ that describes the nonlinear process arising in the system. The last part of the equation describes a time-dependent input signal distributing input information among reservoir nodes. When the reservoir is excited, signals propagating between the nodes give rise to complex dynamics. A linear combination of the reservoir states is read out at certain times and used as an output of the network. Training consists of adjusting the output weights so that the network performs the desired task.

\subsection{Neuromorphic computing with exciton-polaritons}

 %In the hardware polariton neural network realisation, the nonlinear response from the system, triggered by optical signals, recreates an artificial neuron activation function.

 %pioneering idea of harnessing exciton-polaritons in hardware machine learning systems 

Excellent transport properties, together with a strong optical nonlinearity, make exciton-polaritons an ideal platform for realising neuromorphic computing devices. The idea of polariton neuron was first considered in Ref.~\cite{Shelykh_Neurons}. A method for constructing a polariton perceptron based on the superposition of plane waves in a spatially modulated external potential was proposed in~\cite{Ortega_2015}. 

\begin{figure}
\centering
  \includegraphics[width=\textwidth]{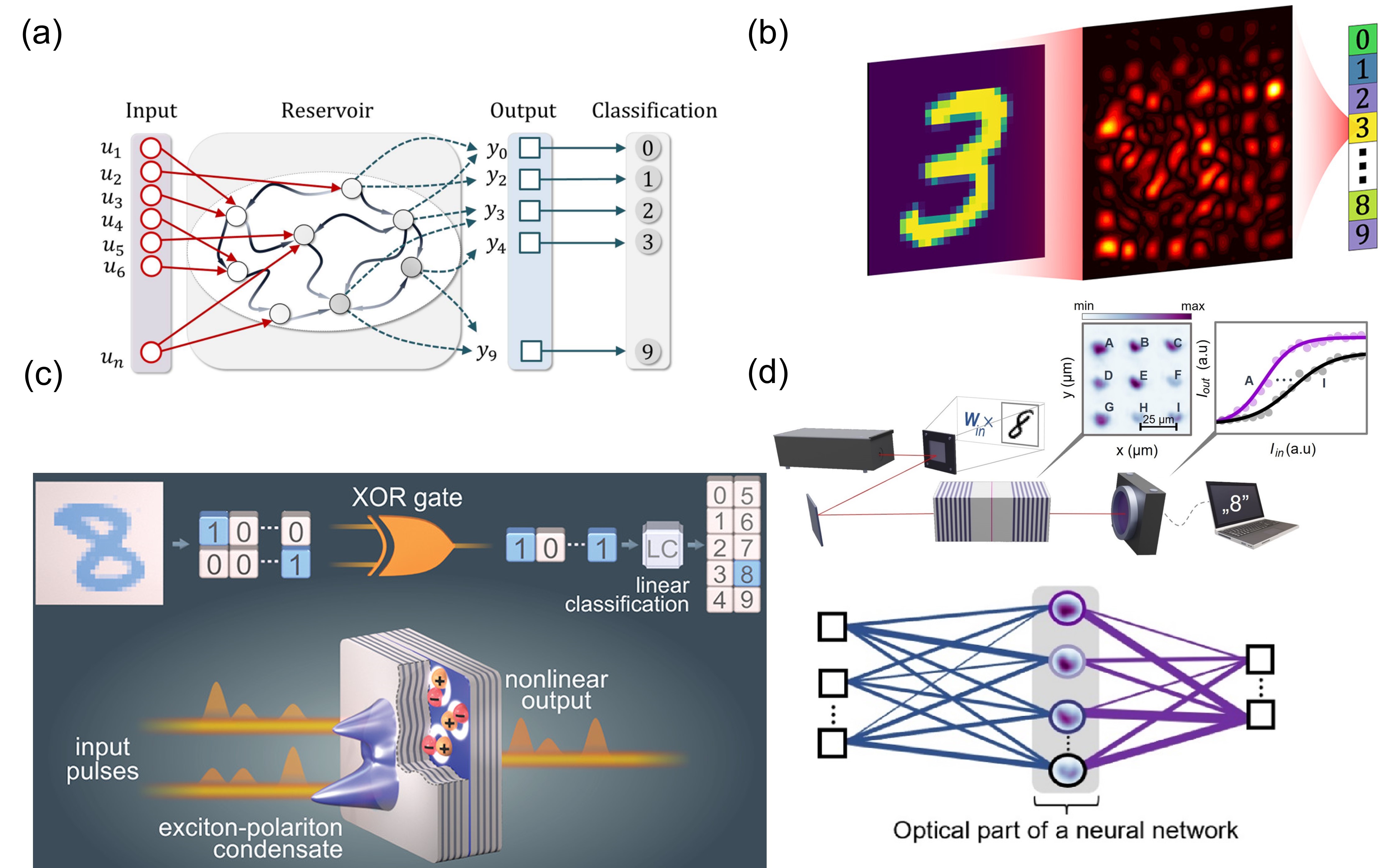}
  \caption{{\bf Polaritons for neuromorphic computing.} (a) Scheme of reservoir computing~\cite{Opala_2019}. (b) State of a polariton reservoir neural network for the input digit "3" from the MNIST dataset~\cite{Ballarini_2020}. (c) Scheme of experimental binarized polariton neural network realization~\cite{Mirek_2021}. (d) Feed-forward exciton polariton neural network~\cite{Opala_2019}.}
\label{fig:fig_2}
\end{figure}

In~\cite{Opala_2019} it was demonstrated that systems described by the complex discrete Ginzburg-Landau equation (CGLE) can be used to realise reservoir neural networks. Complex Ginzburg-Landau equation is one of the fundamental theoretical models describing a wide class of nonlinear wave phenomena in various fields of science. Additionally, CGLE is considered to be the simplest dissipative model describing exciton-polariton condensates~\cite{Bobrovska_2015}. In~\cite{Opala_2019}, a reservoir created by exciton-polariton nodes coupled with each other in a two-dimensional rectangular array was considered. The proposed system is based on the reservoir computing approach, where couplings between neighbouring nodes were fixed, positive, and random, as illustrated in Fig.~\ref{fig:fig_2}(a). In this theoretical proposal, the input was encoded by a sequence of laser pulses. The pulses initiate complex dynamics of reservoir nodes (here, an array of polariton condensates). Dynamics of these nodes mapped input information to a highly nonlinear output space, extracting hidden correlations between input data. The light emitted by the array is collected by a camera, and its intensity is interpreted as an output of the reservoir layer. This output is then multiplied by output weights in software, which is also used for training the network.
The above system was numerically simulated and used to solve machine learning problems such as image classification, nonlinear time series prediction, and speech recognition, achieving high accuracy rates in all cases. The efficiency of the procedure described above results from strong polariton interactions inducing nonlinear dynamics of polariton condensates.  Numerical analysis demonstrated that the optimal working point of such a polariton neural network is close to the condensation threshold. Additionally, it was demonstrated that polariton reservoir neural networks can be used to construct self-correcting networks without the need for error correction algorithms, and implement the function of cascadable Toffoli gates as well as composite circuits~\cite{Xu_2020}.

Shortly after, the first hardware exciton-polariton based neural network was demonstrated~\cite{Ballarini_2020}. This experimental realization  included a lattice of driven-dissipative polariton condensates, forming a physical reservoir. Information was injected in the system in the form of laser pulses, via a spatial light modulator that patterned the laser beam, illuminating a semiconductor microcavity. The input beam was structured as a small square lattice, covering an area of 150 $\mu$m$\times$150 $\mu$m, in which the intensities of lattice nodes encoded the input information.  Optically created polariton condensates subsequently interacted with each other and evolved until reaching a metastable steady state.  Due to the internal reservoir randomness and the interference of polariton nodes, the reservoir output took the form of complex spatially inhomogeneous patterns, as presented  in Fig.~\ref{fig:fig_2}(b). This metastable state was subsequently used as the output of the reservoir layer. Compared to the theoretical proposal~\cite{Opala_2019}, this setup was based only on spatial, not temporal, coding of input data. This experiment demonstrated the efficacy of polariton neural networks and demonstrated a significant enhancement in the accuracy of MNIST (Modified National Institute of Standards and Technology database) pattern recognition task, as compared to linear classification algorithms, and outperforming previous hardware reservoir realizations.  
%The dataset used in this experimental setup comprised 1000 images of ten different classes, which were recognized with an accuracy of 93.1$\%$. 

To achieve accuracy comparable to the most advanced software neural networks, reservoir neural network would require a very high number of internal nodes. On the other hand, reservoir neural networks are following physical system dynamics, which can limit their speed of operation. As was estimated by the authors of Ref.~\cite{Ballarini_2020}, in the ideal case (considering only fundamental physical limitations), polariton systems are able to process the data with speed in the range of THz and with very low energy consumption.  

Subsequent experimental realizations of polariton neural networks operated in the feed-forward architecture, with no connections between nodes in the same layer~\cite{Mirek_2021, Opala_2022}. These networks can process information more accurately, faster, and with lower energy consumption as compared to reservoir neural networks. Experiment~\cite{Mirek_2021} confirmed that polariton neurons can be highly energy efficient if operating in the all-optical scheme. In this setup, it was shown that the realization of a single synaptic operation can cost as little as 16 pJ of optical energy, which is comparable to the cost of operations in the most energy-efficient electronic systems. Therefore, it demonstrates that optics boosted by exciton-polaritons, in comparison to conventional nonlinear optics, can be used for more efficient computations in the nonlinear case. Such efficiency was possible by using highly nonlinear polariton node and applying a time-multiplexed all-optical XOR (exclusive-OR ) logic gate as a neuron of a binarized neural network. Polariton logic gate was realized using two spatially separated localized condensates separated by 2 $\mu$m, excited by off-resonant picosecond laser pulses. 
This binarized neural network encoded $(0)$ and $(1)$ states with low- and high-energy input laser pulses, respectively. Emission from the polariton sample implemented a nonlinear transformation analogous to the operation of a XOR gate. The optical XOR gate took the role of a binary neuron in a single layer of a feed-forward neural network as in  Fig.~\ref{fig:fig_2}(c). In this experiment, only the weights between the hidden and output layers were tuned. Therefore, this neural network can be considered a realization of an extreme learning machine. This neural network can solve the MNIST pattern recognition task with an accuracy of about $96\%$, while keeping low energy consumption. This result is comparable to state-of-the-art implementations of neuromorphic devices.

Polariton neural network based on time-delayed effects in exciton-polariton condensates was reported in~\cite{Mirek_2022}. In this proof-of-principle experiment,  polariton XOR gates were encoded in the time domain. Compared to the work~\cite{Ballarini_2020}, where light-matter quasi-particles were coupled because of the ballistic expansion of the polariton wavefunction, here interaction resulted from the scattering with incoherent reservoir of excitons. Thanks to the memory-like properties of the long-living incoherent excitons, this system could process information temporally on the picosecond time scale. In the same work, based on numerical simulations of the experiment, the authors showed that the binarized XOR polariton neural network could solve speech recognition tasks. Additionally, the work~\cite{Mirek_2022} captured that the oscillatory dynamics of exciton-polaritons observed under incoherent pulse excitation possesses many analogies to biological neurons. 

This analogy was explored in Ref.~\cite{Tyszka_2023}, where authors showed the analogy between the polariton condensate evolution and the Leaky Integrate-and-Fire (LIF) model of a spiking neuron. The LIF model was proposed to mimic the dynamics of a biological neuron in the brain. This neuron model assumes temporal integration of all input spikes incoming to the neuron, resulting in the accumulation of the so-called membrane potential. When the membrane potential overcomes certain threshold, the neuron emits an output spike. Observation of nonresonantly excited polariton quantum fluids where long-lived excitons are present and the appearance of stimulated scattering processes between the polariton condensate and the incoherent reservoir suggest the existence of similarities between polariton condensates and the LIF model. The analysis presented in~\cite{Tyszka_2023} indicated that ultrafast spike-like signals of the polariton field could be used for information processing with energy efficiency below one pJ per spike.

It should be underlined that the polariton neural networks presented in~\cite{Ballarini_2020, Mirek_2021, Mirek_2022} worked in the reservoir or ELM approach. In both cases, only the weights between the last hidden layer and the output layer were adjusted, as explained in the previous sections. These approximations work well only when the number of neurons in the hidden layer is large enough, which is an obstacle in the case of complex tasks of practical relevance. To provide high accuracy in the case of a relatively small number of neurons, in the work~\cite{Opala_2022}, a feed-forward neural network architecture utilizing the backpropagation algorithm was implemented, see  Fig.~\ref{fig:fig_2}(d). The application of the backpropagation algorithm resulted in the reduction of the number of neurons from $10^5$ in work ~\cite{Mirek_2021} to $90$, while keeping the same recognition rate. In this experimental realization, a $3 \times 3$ polariton lattice was realized using a spatial light modulator. Light intensity needed to control a single polariton neuron with a radius of approximately 6 $\mu$m was only 9 $\mu W/\mu m^2$. In contrast to previous works, weights corresponding to both the input layer and the output layer were optimized in the training phase. This was possible thanks to the experimental characterisation of the optical response of each polariton mode together with the in-silico training method. The result of multiplication of input data by input weights was encoded directly by the SLM, where output weights were implemented using post-processing performed in software.

Although the experiments described above were performed in systems where large part of computations was performed in software, the true potential of exciton-polariton neural networks should be revealed in all-optical networks~\cite{Matuszewski_2021}. In this case, thanks to the nonlinearity orders of magnitude stronger than in other optical systems, polaritons should be able to form a basis of hardware neural networks with exceptionally high energy efficiency. This is due to the possibility of performing both efficient linear and nonlinear operations in a single system, without the need for conversion from optical to electronic signals and vice versa. Theoretical estimates indicate that efficiency of such systems may reach attojoules per a single neuron activation, which is orders of magnitude higher than in specialized electronic systems such as tensor processing units (TPUs)~\cite{Matuszewski_2021,jouppi2017datacenter}. However, to achieve this goal it is necessary to design and construct a large-scale all-optical system working at room temperature, with the capability of signal regeneration between layers, which is currently a significant technological challenge.

\section{{Polariton condensates for optimization}}\label{sec:optimization}

In recent years, exciton-polariton condensates have been studied as a platform for complex optimization problems such as determining ground states of spin models~\cite{berloff2017realizing} or various NP-hard (non-deterministic polynomial-time hard) problems~\cite{kyriienko2019probabilistic}. In principle, specialized optimizers can efficiently solve problems that are intractable for classical computers working according to the von Neumann architecture.
This results from the large number of degrees of freedom and the computational complexity. %Examples include non-deterministic polynomial-time hard problems and simulations of many-body systems.
The concept underlying polariton simulators involves reformulating complex computational problems as spin systems, described, for example, by Ising, Heisenberg, or $XY$ models~\cite{stroev2023renaissance}. In this context, the classical spin vector can be interpreted as the phase of a polariton condensate, thus enabling the optimization of spin Hamiltonians in a polariton system converging to a global ground state, in which the energy of the spin model is minimized.

The interest in using exciton-polaritons for optimization was sparked by the experiment from 2017~\cite{berloff2017realizing}, which demonstrated that lattices of polariton condensates excited by nonresonant laser beams can be used to determine energy minima of $XY$ models. Pumping beams can be patterned using light modulators to an arbitrary spatial configuration, which results in high flexibility and reconfigurability. Moreover, connections between polariton nodes can be engineered, which allowed to realize various phases including ferromagnetic, antiferromagnetic and spin-frustrated configurations. Theoretical considerations showed that a polariton system at the condensation threshold has the ability to find the global ground state in the bottom-up approach, akin to the coherent Ising machine, which can greatly facilitate the optimization process~\cite{lagoudakis2017polariton}. Moreover, the same system can be used above threshold to study spin synchronization and dynamical phase transitions.

Subsequent theoretical works demonstrated that specific configurations of polariton networks can be used to simulate other spin systems including Ising and n-state planar Potts models~\cite{kalinin2018networks}, k-local Hamiltonians~\cite{stroev2021discrete}, and coupled oscillator models including Kuramoto and Stuart-Landau models~\cite{kalinin2019polaritonic}. Moreover, they can can exhibit exotic phases such as the spin-liquid state~\cite{kalinin2018exotic}. The problem of a priori unknown occupations of polariton nodes, which are complex functions of external pumping, was also investigated~\cite{kalinin2018networks}. While in principle it prohibits solving NP-hard problems, it was shown that it can be remedied by an appropriate optical feedback scheme. Recently, it was argued that polariton networks belong to the class of Andronov-Hopf oscillator networks realizing coherent $XY$-Ising machines~\cite{syed2023physics}. It was also shown numerically and experimentally that coherent polariton networks can be used for solving the max-3-cut problem~\cite{harrison2022solving} which can be applied for image segmentation. 

Alternative methods for optimization using exciton-polaritons that have been proposed include all-to-all coupling between different spatial eigenmodes of a trapped condensate~\cite{sigurdsson2018all}. In this physical setting, the system seeks the maximum of gain that corresponds to the highest eigenvalue of the matrix of weights. In Ref.~\cite{kyriienko2019probabilistic}, a method for solving NP-hard problems with polariton lattice systems in the bistable regime was put forward. This setting allows for nonlinear optimization of Ising models with an effective transverse field and all-to-all coupling, spin glass, graph partitioning and the knapsack problem.

\section{Quantum computing and other approaches to computation} \label{sec:other}

In addition to the computation schemes described above, several works proposed that exciton-polariton systems can be used for alternative computing schemes. It was shown that lattices of exciton-polariton nodes with certain spin dependent-pumping can be used to realize a physical model of artificial life~\cite{banerjee2020artificial}, which is known to be able to perform universal computation. Several other work investigated the possibility to realize quantum computing with exciton-polaritons. It was proposed that engineering parametric interaction between polariton modes can be used to construct controlled-NOT quantum gates~\cite{kyriienko2016exciton}. Another way to realize controlled-NOT gate using spin control in a patterned polariton system was proposed in Ref.~\cite{solnyshkov2015all}. Yet another work introduced an universal set of one- and two-qubit quantum gates with polaritons in quantum dots~\cite{puri2017universal}. It was also demonstrated that quantum fluctuations on top of polariton condensates can be harnessed to realize qubits, which can be controlled externally by applied laser pulses~\cite{ghosh2020quantum}. Another proposal for a polariton qubit involves a condensate in a split-ring resonator~\cite{xue2021split}.

In the context of neural networks, a promising possibility is the use of exciton-polaritons for quantum reservoir computing and processing~\cite{Ghosh_2019_QRP,Ghosh_2021_QNC}. In contrast to classical reservoir computing networks, here the reservoir is based on quantum nodes (eg.~qubits), and performs a transformation of a many-body quantum state of the system. This approach can in principle lead to tremendous advantages with respect to the classical case, due to the greatly increased dimensionality of the space of allowed states~\cite{fujii2017harnessing,Ghosh_2021_QNC}. This in turn can lead to superpolynomial quantum enhancement of neuromorphic computing~\cite{xu2021superpolynomial}. Moreover, it allows to use quantum networks for quantum tasks, which is obviously not possible with classical neural networks. These tasks include quantum state preparation~\cite{Ghosh_2019_QNP}, quantum state reconstruction~\cite{Ghosh_2021_RCS}, reproducing quantum circuit operation~\cite{ghosh2021realising} and beating the classical limit of phase estimation~\cite{krisnanda2022phase}. The role of the nonlinearity in quantum reservoir networks was also considered~\cite{Xu_2023}.

\section{Conclusions}

The discovery of quantum fluids of light in exciton-polariton microcavities has sparked a renewed interest in implementing computations solely with light. Strong quantum coupling of light and matter facilitates this goal, making it easier to realize basic  nonlinear elements such as switches, gates and transistors. At the same time, wide range of methods of manipulating  physical properties of polaritons and their effective potential landscapes, including the effective mass, and their non-Hermitian interactions, opened the way to precise control over their dynamics. An important challenge for polariton information processing is scaling up these basic elements to fully-functional circuits, which requires fulfilling basic conditions of cascadability in a large-scale system. 

From this point of view, non-von-Neumann architectures such as neuromorphic computing and hardware implemented optimization are interesting avenues that could make optical computing competitive with the established electronic systems. Up to now, systems consisting of between a few to several tens of computing units have been implemented~\cite{Ballarini_2020,lagoudakis2017polariton}, which is not enough for solving complex, practically relevant problems. For neuromorphic systems, the issue of energy efficiency is probably the most important~\cite{mehonic2022brain}. According to theoretical estimates, efficiencies orders of magnitude higher than in electronic systems could be achieved~\cite{Matuszewski_2021}. However, this requires the implementation of a strong polariton nonlinearity at room temperature, with limited losses and at high data rates. In the case of optimization by polariton networks, the possibility to solve hard problems gives an advantage over any standard classical computing system at a sufficiently large scale. Therefore, scaling polariton optimization networks, while maintaining sufficiently high accuracy, appears to be the most important challenge for future experiments.

%~\cite{Ortega_2015,Berloff_2017, Opala_2019, Ghosh_2019_QNP,Ghosh_2019_QRP, Ballarini_2020,Xu_2020,Mirek_2021, Ghosh_2021_RCS, Ghosh_2021_QNC,Topfer_21, Matuszewski_2021,Mirek_2022, Opala_2022,Tyszka_2023,Xu_2023}

{\bf Funding.} We acknowledge support from the National Science Center, Poland grants 2020/37/B/ST3/01657 and 2021/43/B/ST3/00752.

{\bf Disclosures.} The authors declare no conflicts of interest.

{\bf Data availability.} No data were generated or analyzed in the presented research.

%%%%%%%%%% If using BibTeX:
\bibliography{omex}

\end{document}